# Design And Fabrication of Condenser Microphone Using Wafer Transfer And Micro-electroplating Technique

Zhen-Zhun Shu[1]  Ming-Li Ke[1]  Guan-Wei Chen[1]  Ray-Hua Horng[1]
Chao-Chih Chang[2]  Jean-Yih Tsai[2]  Chung-Ching Lai[2]  Ji-Liang Chen[2]
Institute of Precision Engineering, National Chung Hsing University, Taichung, Taiwan, R.O.C. [1]
Taiwan Carol Electronics Co. Ltd., NO. 202, Tung Kuang Road, Taichung, Taiwan, R.O.C. [2]

*Abstract- A novel fabrication process, which uses wafer transfer and micro-electroplating technique, has been proposed and tested. In this paper, the effects of the diaphragm thickness and stress, the air-gap thickness, and the area ratio of acoustic holes to backplate on the sensitivity of the condenser microphone have been demonstrated since the performance of the microphone depends on these parameters. The microphone diaphragm has been designed with a diameter and thickness of 1.9 mm and 0.6 μm, respectively, an air-gap thickness of 10 μm, and a 24% area ratio of acoustic holes to backplate. To obtain a lower initial stress, the material used for the diaphragm is polyimide. The measured sensitivities of the microphone at the bias voltages of 24 V and 12 V are -45.3 and -50.2 dB/Pa (at 1 kHz), respectively. The fabricated microphone shows a flat frequency response extending to 20 kHz.*

## I. INTRODUCTION

Microphones are electro-acoustic transducers that convert acoustical energy into electric energy. At present, electro-acoustic transducers cover a wide field of applications in cellular phones, digital cameras, hearing aids and so on. There are few different designs of microphones. Most of the silicon microphones are condenser microphones because of their high sensitivity, low noise level, and flat frequency responses in a wide bandwidth [1,2]. Condenser microphones consist of two parallel plates, namely a flexible diaphragm and a rigid backplate. During the last decade, the miniature microphones based on silicon micromachining techniques were the subject of research and development [3]. In general, for the micro-electro-mechanical system (MEMS) microphone applications, mechanical structures are fabricated on silicon wafers using thin film deposition, lithography and etching techniques [4,5]. Coupled with matured silicon micromachining and MEMS techniques, the size scale of microphones can be decreased efficiently. In fact, lowering of the sensitivity is obtained by condenser microphone miniaturization. To improve its mechanical sensitivity, a corrugated diaphragm and several complex processes for making silicon condenser microphones have been reported [6-9]. Although these researches achieved high mechanical sensitivities, the extra processes will increase the production cost. To overcome these disadvantages, the condenser microphone fabricated by the wafer transfer and micro-electroplating technique wherein the substrate can also be recycled. This paper describes the new fabrication process and design as well as the acoustic characterization of the microphone.

## II. Microphone Design

Condenser microphones consist of two parallel plates which form a variable gap capacitor. Figure 1 shows the basic structure of the condenser microphone. The upper plate is a thin and flexible diaphragm which can be made of polyimide, silicon nitride, or polysilicon. The lower plate is a thick and fixed backplate. The open-circuit sensitivity $S_o$ of a condenser microphone is defined by the product of the diaphragm mechanical sensitivity $S_m$ and the electrical sensitivity $S_e$:

$$S_o = S_m S_e \qquad (1)$$

### A. Diaphragm

The mechanical sensitivity of the microphone can be expressed as

$$S_m = \frac{dw}{dP} \qquad (2)$$

where w is the deflection of the diaphragm, and P is the pressure acting on the diaphragm. At low frequencies, the mechanical sensitivity is approximately given by [10]

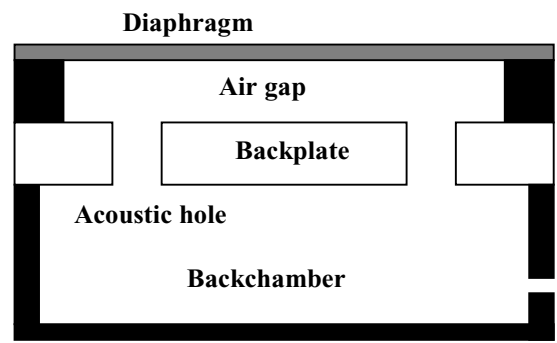

Fig. 1. Structure of the condenser microphone.

$$S_m = \frac{A}{8\pi\sigma h_d} \quad (3)$$

where $A$, $\sigma$ and $h_d$ are the diaphragm's area, tension stress and thickness, respectively. Thus, choosing the diaphragm material to obtain a lower initial stress will result to a higher mechanical sensitivity [11], on the other hand, microphones with thinner diaphragm have higher mechanical sensitivity. Mechanical sensitivity is obtained by measuring the central deflection of the diaphragm under a given applied pressure as shown in Fig. 2. It can be seen that the mechanical sensitivity increases when the diaphragm thickness is reduced. In this study, the design thickness and diameter are 0.6 µm, 1.9 mm, respectively, with polyimide as the diaphragm material.

*B. Air gap*

The bias voltage ($V_b$) applied to the condenser microphone provides an electric field in the air gap between the diaphragm and the backplate. The sound pressure variations cause a movement in the microphone diaphragm, resulting to a change in the voltage across the air gap. The electrical sensitivity of the microphone can be expressed as

$$S_e = \frac{\Delta V_b}{\Delta d} = \frac{V_b}{d} \quad (4)$$

where $d$ is the air gap thickness. From the above equation, the electrical sensitivity is proportional to the bias voltage and inversely proportional to the air gap thickness. Theoretically, a higher voltage bias can increase the electrical sensitivity. However, the bias voltage is limited by the pull-in voltage ($V_p$) of the microphone ($V_b < V_p$). The pull-in voltage for a circular diaphragm can be expressed as [12]

$$V_p = \sqrt{\frac{8}{27}\frac{d^3}{\varepsilon_0 S_m}} \quad (5)$$

where $\varepsilon_0$ is the permittivity in air.

Combining equations (1), (4) and (5), the open-circuit sensitivity of the microphone would be

$$S_o < \sqrt{\frac{8dS_m}{27\varepsilon_0}} \quad (6)$$

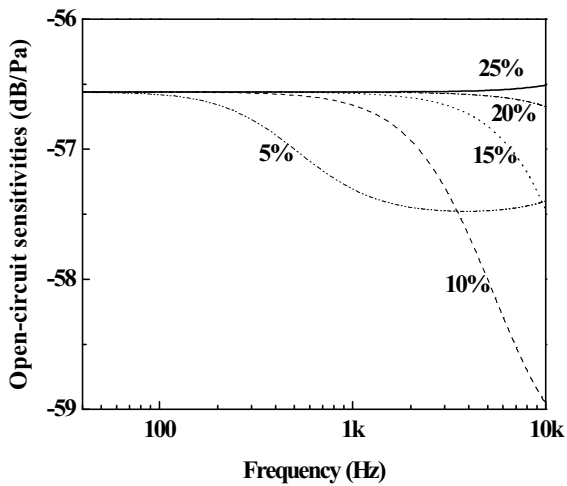

Fig. 3. Frequency response as a function of the area fraction of the acoustic holes.

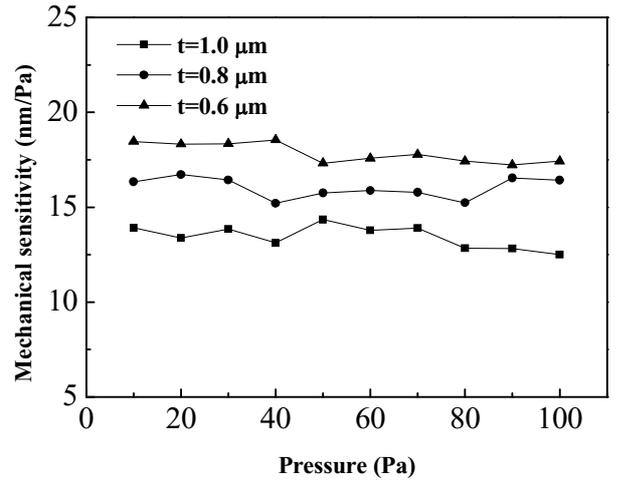

Fig. 2. Mechanical sensitivity at different diaphragm thicknesses.

At this pull-in voltage, the diaphragm touches the blackplate and causes an electrical short. Although increasing the air gap thickness results to a higher open-circuit sensitivity and pull-in voltage, it also reduces the air resistance of the microphone. Nevertheless, the air gap thickness cannot be increased indefinitely since the capacitance between the diaphragm and the backplate will decrease. Hence, the air gap thickness was set to 10 μm in this study.

*C. Backplate*

The air resistance of the air gap ($R_a$) and the backplate hole ($R_h$) can be expressed as [13]

$$R_a = \frac{12\nu A_{eff}}{\pi d^3 n}(\frac{A}{2} - \frac{A^2}{8} - \frac{\ln A}{4} - \frac{3}{8}) \quad (7)$$

and

$$R_h = \frac{8\nu t A_{eff}}{n\pi r^4} \quad (8)$$

respectively, where $\nu$ is the viscosity of air, $A_{eff}$ is the area of the backplate, $A$ is the ratio of the total acoustic holes to the total backplate area, $n$ is the density of acoustic holes, $t$ is the backplate thickness, and $r$ is the radius of the backplate hole. The air damping and frequency response are controlled by the acoustic holes of the backplate. Figure 3 illustrates the effect of the area fraction of acoustic holes on the frequency response. It can be seen that the cut of frequency and sensitivity increases when the area fraction of acoustic holes is raised. Thus, when the area fraction of the acoustic holes is increased to 25%, the microphone sensitivity is raised. In this research, about 24% area fraction of 80 μm × 80 μm acoustic holes on a backplate that is 1.9 mm in diameter and 100 μm thick was designed.

The performance of the microphone depends on the size of diaphragm, air gap thickness, the area ratio of acoustic holes to backplate, and so on. To optimize the sensitivity, the parameters used for the proposed MEMS condenser microphone are shown in Table 1.

Table 1
Design parameters of the MEMS condenser microphone.

| Parameters | Value |
|---|---|
| Diaphragm material | Polyimide |
| Diaphragm diameter | 1.9 mm |
| Diaphragm thickness | 0.6 µm |
| Air gap | 10 µm |
| Acoustic hole size | 80 µm × 80 µm |
| Backplate thickness | 100 µm |
| Bias voltage | 12 V |

### III. Fabrication Process

*A. Diaphragm*

The diaphragm and the backplate were fabricated in different wafers. Figure 4 (a) shows the fabrication process of the diaphragm. About 0.5 µm thick of $SiO_2$ film was deposited on the p-type (1 0 0) wafer by LPCVD. Then, a polyimide layer is spin-coated on the oxide layer and patterned by photolithographic techniques. The 80 µm thick nickel frame was formed by electroplating after the electrode material (Cr/Au) was deposited on the diaphragm. Finally, the diaphragm chip was lifted off by the BOE.

*B. Backplate*

The microphone backplate was also fabricated by wafer transfer technique. Figure 4 (b) illustrates the backplate process which begins with $SiO_2$ deposition by LPCVD, and followed by the deposition of 200 nm of Cr/Au seed layer on the $SiO_2$ film using the evaporation process. This seed layer is patterned and etched to form the square-shaped acoustic holes.

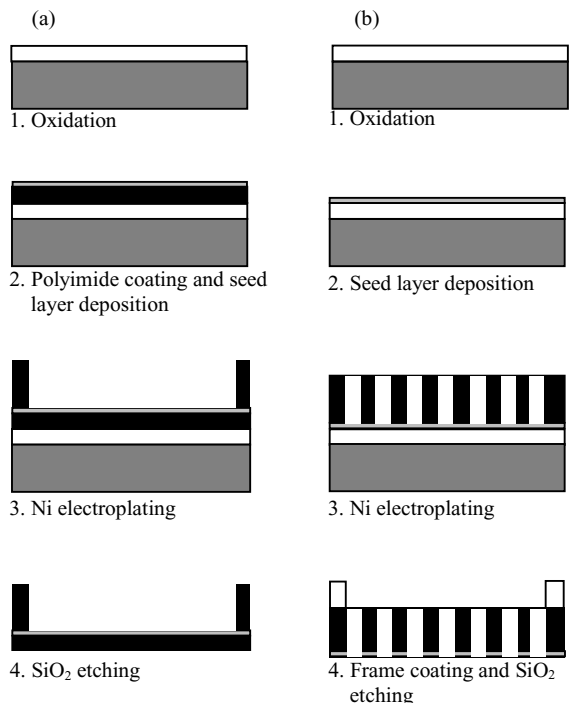

Fig. 4. Process flow for the condenser microphone: (a) diaphragm; (b) backplate.

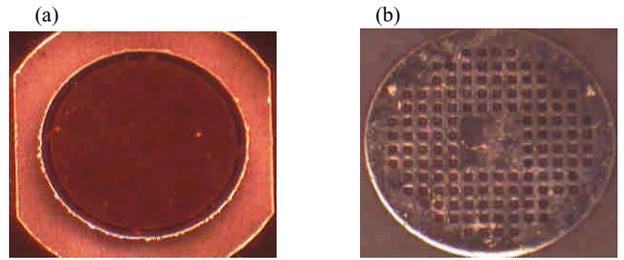

Fig. 5. Photographs of the condenser microphone: (a) diaphragm; (b) backplate.

The backplate structure was fabricated with nickel using the micro-electroplating technique. A 10 µm thick SU8 spacer is patterned and used to define the air gap between the diaphragm and the backplate. The microphone components were separated from the substrate by the BOE.

*C. Prototype microphone*

Figure 5(a) and (b) show the fabricated diaphragm and backplate chips, respectively. The two components of the condenser microphone are mechanically clamped together. To make the prototype condenser microphone, the chip is packaged in a metal housing with a preamplifier circuit as shown in Figure 6.

### IV. Results and Discussion

*A. Stress of the diaphragm*

The mechanical sensitivity $S_m$ values obtained are 13, 15, and 17 nm/Pa for the diaphragm thickness values of 1, 0.8, and 0.6 nm, respectively. Substituting the diaphragm's area, thickness, and mechanical sensitivity values into equation (3), the calculated average internal stress of the diaphragms are 8.68, 7.52, and 6.64 MPa. According to equation (3), mechanical sensitivity can be increased by thinning the diaphragm and decreasing the internal stress. Stress is a very important property of a microphone diaphragm. The thermal process should also be controlled in order to decrease the internal tensile stress of the diaphragm. Several materials were tried in the fabrication of diaphragm, but only polyimide has the advantage of having a low temperature fabrication process.

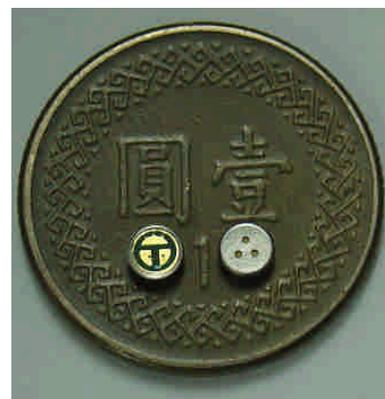

Fig. 6. A packaged condenser microphone on a coin.

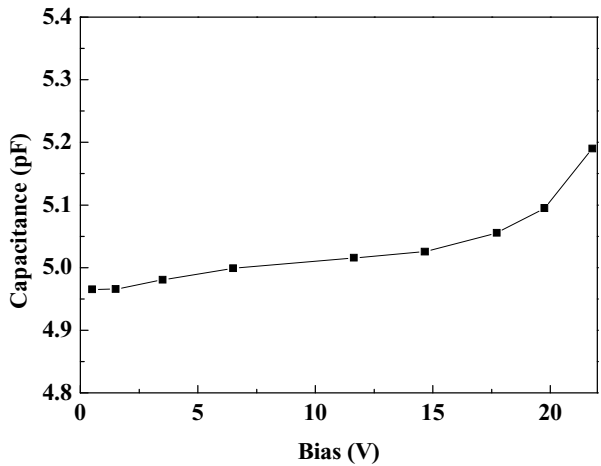

Fig. 7. The capacitance characteristics for a microphone at different voltage biases

*B. Electrical measurements*

Using the mechanical sensitivity value of 17 nm/Pa for the diaphragm thickness of 0.6 μm (Fig. 4), the pull-in voltage is calculated to be about 44 V. Using an impedance analyzer (HP-4284), the capacitances of the packaged microphone were determined. Figure 7 plots the capacitance against the voltage bias without an applied static pressure. The capacitance increased as the voltage bias increased because the diaphragm was pulled towards the backplate. It is clear that the microphone chip operates stably when the voltage bias was less than 25 V.

Figure 8 shows the relationship between capacitance and static pressure at two different voltage biases. The complete set-up for measuring the capacitance change involves a pressure vessel. When the static pressure is varied with a pressurization cylinder, the diaphragm of the condenser microphone attached to the backplate moves and results to a change in capacitance. The capacitance increased linearly with pressure up to 100 Pa. The capacitance can also be increased by increasing the voltage bias due to the decrease of the air gap. Although, it can raise the sensitivity of the condenser microphone, its disadvantage is that the band of flat frequency response is reduced because the resonance frequency increases as the square root of the mechanical stiffness.

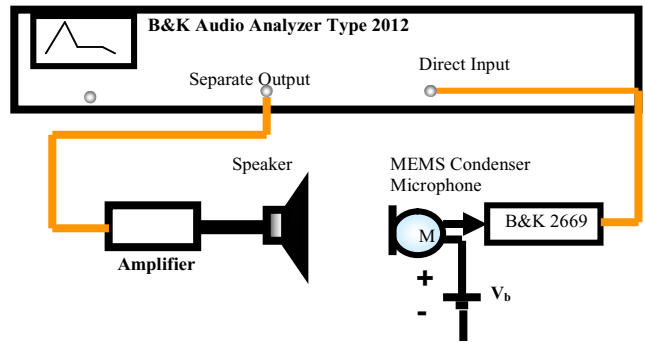

Fig. 9. Schematic diagram of the acoustic response measuring system.

*C. Frequency response*

In order to test the sensitivity of the microphone, the device was placed in an anechoic room as shown in Fig. 9. The fabricated microphone is fitted with a preamplifier (B&K 2669), and the acoustic response is measured using an audio analyzer (B&K 2012). Voltage biases of 12 V and 24 V were used in the measurement.

The frequency response of the prototype microphone is shown in Figure 10. The voltage biases of 12 V and 24 V showed that the sensitivities of the condenser microphone were at -50.2 and -45.3 dB/Pa (at 1 kHz), respectively. Obviously, the sensitivity of the microphone increases with the voltage bias because the electrical sensitivity increases. However, a flat frequency response was obtained over the audio frequency range.

V. Conclusion

The sensitivity of the MEMS condenser microphone can be increased by increasing the mechanical and electrical sensitivities. Any attempt to optimize it will likely result into a compromise somewhere. For example, increasing the diaphragm area and voltage bias, as well as decreasing the diaphragm stress and thickness will raise the sensitivity, but the band of the flat frequency response is reduced. In this paper, the design and fabrication of a high sensitivity microphone is presented. The condenser microphone is fabricated by the MEMS technology through the wafer transfer and micro-electroplating techniques in order to avoid many extra processes. Based on the simulation results, when the area fraction of the acoustic holes is increased to 25%, outstanding microphone sensitivity could be obtained. The mechanical sensitivity at the diaphragm thickness of 0.6 μm is 17 nm/Pa. Moreover, a sensitivity of up to -50.2 dB/Pa was measured for the fabricated microphone with a diaphragm diameter of 1.9 mm and a thickness of 0.6 μm, using a voltage bias of 12 V. Moreover, the fabricated microphone shows a flat frequency response in the audio frequency range.

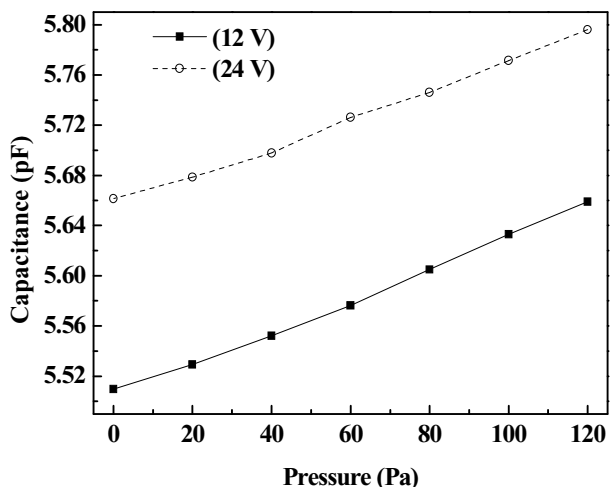

Fig. 8. Pressure versus capacitance characteristics of the microphone.

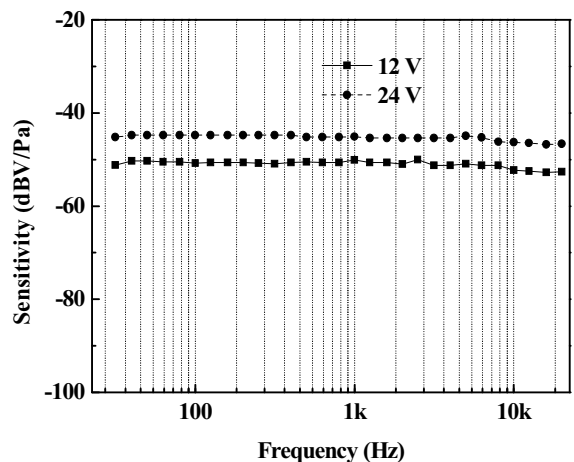

Fig. 10. Frequency response of the prototype microphone.